\documentclass[aps,prl,superscriptaddress,reprint]{revtex4-1}
\usepackage{amsmath}
\usepackage{amssymb}
\usepackage{graphicx}
\usepackage[colorlinks,urlcolor=blue]{hyperref}
\usepackage{url}
\usepackage{color,xcolor}
\usepackage{ulem}
\usepackage{float}
\usepackage{epstopdf}
\begin{document}

\title{$J = 0$ nonmagnetic insulating state in K$_2$Os$X_6$ (X = F, Cl and Br)}
\author{Yang Zhang}
\author{Ling-Fang Lin}
\affiliation{Department of Physics and Astronomy, University of Tennessee, Knoxville, TN 37996, USA}
\author{Adriana Moreo}
\author{Elbio Dagotto}
\affiliation{Department of Physics and Astronomy, University of Tennessee, Knoxville, TN 37996, USA}
\affiliation{Materials Science and Technology Division, Oak Ridge National Laboratory, Oak Ridge, TN 37831, USA}
\date{\today}

\begin{abstract}
In $4d/5d$ transition-metal systems, many interesting physical properties arise from the interplay of bandwidth, electronic correlations, and spin-orbit interactions. Here, using {\it ab initio} density functional theory, we systematically study the double-perovskite-like system K$_2$Os$X_6$ (X = F, Cl, and Br) with a $5d^4$ electronic configuration. Our main result is that the $J = 0$ nonmagnetic insulating state develops in this system, induced by strong spin-orbit coupling (SOC). Specifically, the well-separated Os$X_6$ octahedra lead to the cubic crystal-field limit and result in dramatically decreasing hoppings among nearest neighbor Os-Os sites. In this case, the three degenerate $t_{2g}$ orbitals are reconstructed into two ``effective'' $j_{\rm eff}$ ($j_{\rm eff} = 1/2$ and $j_{\rm eff} = 3/2$ states) states separated by the strong SOC, opening a gap with four electrons occupying the $j_{\rm eff} = 3/2$ orbitals. Furthermore, the hybridization between the Os $5d$ orbitals and the $X$ ($X$ = F, Cl, and Br) $p$ orbitals increases from F to Br, leading the electrons in K$_2$OsF$_6$ to be more localized than in K$_2$OsCl$_6$ and K$_2$OsBr$_6$, resulting in a smaller bandwidth for K$_2$OsF$_6$ than in the Cl- or Br-cases. Our results provide guidance to experimentalists
and theorists working on this interesting family of osmium halides.
\end{abstract}

\maketitle
In the past decade, material systems with $4d$ or $5d$ transition-metal (TM) atoms, such as iridium and osmion, have attracted growing attention due to the exotic physical phenomena induced by the strong spin-orbit coupling (SOC)~\cite{SOC,Rau:arcm,Cao:rpp,Khomskii:cr,Takayama:jpsj}. Compared to the nearly negligible SOC in $3d$ atoms, the strength of the SOC $\lambda$ is substantially enhanced in those $4d/5d$ systems, leading to comparable values between $\lambda$, the hopping parameter $t$, the Hubbard repulsion $U$, and the Hund coupling $J_H$, resulting in several intriguing electronic phases arising from their competition, such as topological phases~\cite{Xiao:nc,Hasan:rmp,Wan:prb,Yan:arcmp}, the orbital-selective Peierls phase~\cite{Streltsovt:prb14,Zhang:ossp}, ``spin-orbit Mott'' insulating state~\cite{Kim:prl08,Zhang:prb22}, Rashba-like splitting~\cite{Bruyer:prb16,Zhang:prb20}, the anomalous Hall effect~\cite{Mohanta:prb,Yoo:nc}, and quantum spin liquid ground states~\cite{Kitaev:aop,Takagi:nrp}.

\begin{figure}
\centering
\includegraphics[width=0.48\textwidth]{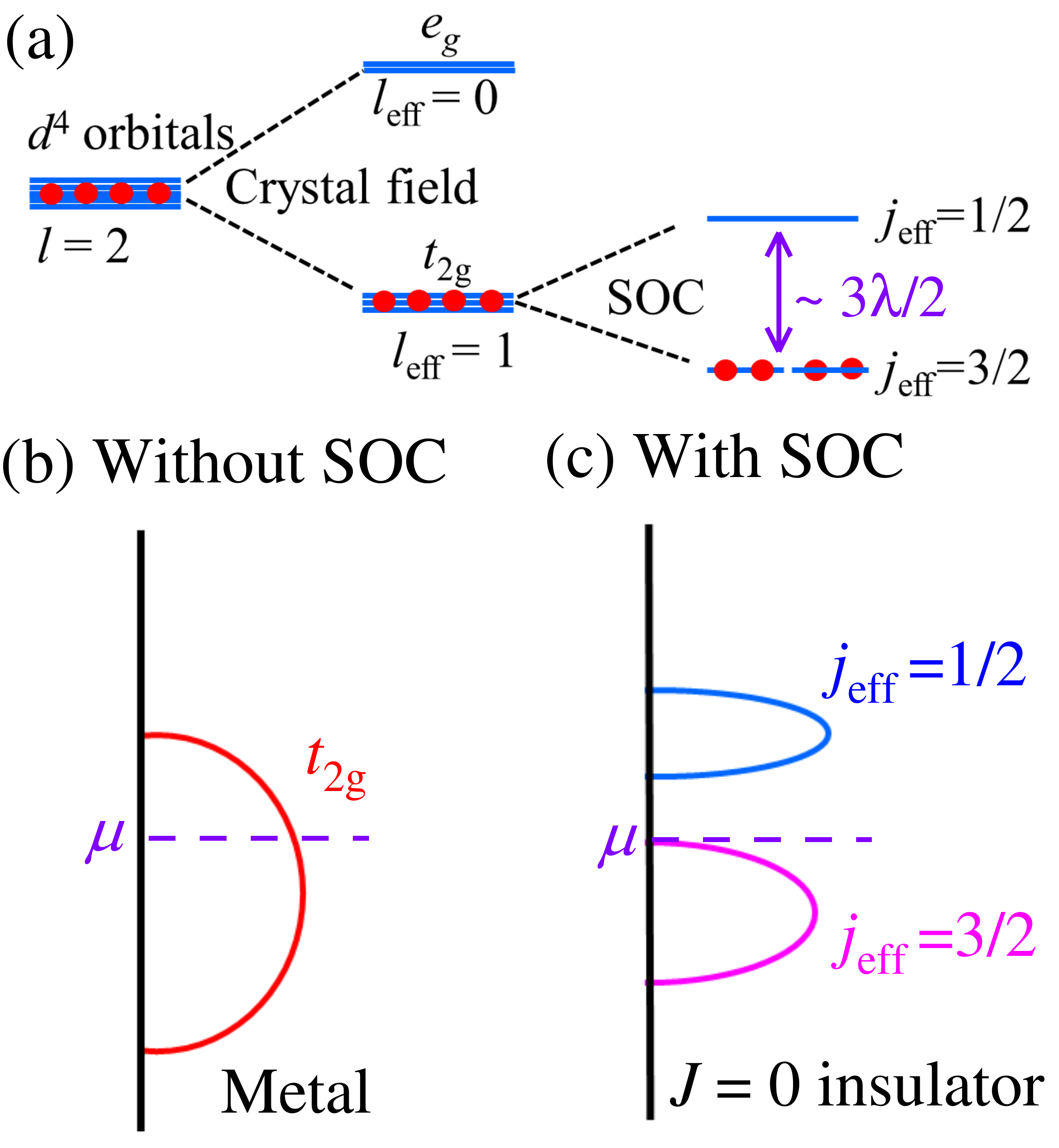}
\caption{(a) Schematic energy splitting of the $d^4$ electronic configuration with strong SOC under a cubic crystal field. Here, only the Hund rule is considered. Note the $j_{\rm eff} =3/2$ state has degeneracy four with $m = \pm  1/2$ and $m = \pm 3/2$. (b) The $t_{\rm 2g}$ orbitals induce a metallic phase in a $d^4$ system without SOC and Hubbard $U$~\cite{d4}. (c) Without Hubbard $U$, a $J = 0$ NM insulator is obtained in the $d^4$ system induced by SOC. Here, $\mu$ is the chemical potential.}
\label{Fig1}
\end{figure}

In an octahedral $d^4$ system, under a cubic crystal-field condition, the five degenerate $d$ orbitals split into two sets of bands (the higher $e_{g}$ and lower $t_{2g}$ bands) separated by a large crystal-field splitting energy $\Delta$ ($\sim 10$ Dq)~\cite{Dq} [Fig.~\ref{Fig1}(a)]. Due to the competition among $\Delta$, $J_H$ and $\lambda$, many electronic states are possible for a $d^4$ system, such as $S = 2$, $S = 1$, and $J =0$ states. For the $4d/5d$ system with $d^4$ configuration, such as Ru$^{4+}$, Os$^{4+}$, and Ir$^{5+}$, the four electrons will occupy the three degenerate $t_{2g}$ orbitals, leading to a metallic state [Fig.~\ref{Fig1}(b)] because the $t_{\rm 2g}$ orbitals are not fully occupied (four electrons in three $t_{\rm 2g}$ orbitals). Then, by introducing the SOC effect, the three $t_{2g}$ orbitals reconstruct into two ``effective'' $j_{\rm eff}$ ($j_{\rm eff} = 1/2$ and $j_{\rm eff} = 3/2$) states separated by the SOC [Fig.~\ref{Fig1}(a)]. Returning to the case $U=0$, and considering the SOC with $\lambda \textgreater 0$, this system with a $d^4$ electronic configuration is expected to be a nonmagnetic (NM) insulator made of local two-hole $J = 0$ singlets~\cite{Khaliullin:prl13,Meetei:prb15}, where the band gap is opened by the strong SOC between the $j_{\rm eff} = 1/2$ and $j_{\rm eff} = 3/2$ states [Fig.~\ref{Fig1}(c)].

Recently, neutron scattering experiments revealed a soft longitudinal magnon mode in Ca$_2$RuO$_4$ with the $4d^4$ configuration, which has been considered as a hallmark of a $J = 0$ singlet state quantum phase transition~\cite{Jain:np,Souliou:prl}. In theory, this could lead to an exotic magnetically ordered state induced by the condensation of spin-orbit excitons~\cite{Khaliullin:prl13,Meetei:prb15,Svoboda:prb17,Kaushal:prb17,Kim:prb17,Sato:prb19,Kaushal:prb20,Kaushal:prb21,Strobel:prb21}. Most searches for the $J = 0$ state have mainly focused on the $5d^4$ iridates systems. However, they display the $S = 1$ magnetic ground state or weak moments, instead of the NM insulator with the $J = 0$ singlets, such as the $5d^4$ iridates Ba$_3$YIr$_2$O$_9$~\cite{Dey:prb12} and Ba$_3$ZnIr$_2$O$_9$~\cite{Nag:prl16,Nag:prl19}. Furthermore, the $J = 0$ state is still under debate for some double perovskite iridates with quite small noncubic crystal-field effect, such as Sr$_2$YIrO$_6$~\cite{Cao:prl14,Bhowal:prb15} and Ba$_2$YIrO$_6$ ~\cite{Terzic:prb17,Fuchs:prl18}. Hence, the $J = 0$ state is still rare in real materials.

In general, the $4d/5d$ orbitals are much more spatially extended than the $3d$ orbitals, leading to an enhanced hopping $t$ in the $4d/5d$ case, corresponding to a large bandwidth $W$ in these systems (with the hopping $t$ providing the scale). In this case, the large bandwidth $W$ would induce the breakdown of the $J = 0$ singlet state~\cite{Bhowal:prb15,Zhang:apl,Zhang:prb22-2}, where an $S = 1$ state is obtained because $W \gg$ $\lambda$. Furthermore, the $J = 0$ singlet ground state could also be suppressed by the large crystal-field splitting energy (between $d_{xz/yz}$ and $d_{xy}$ orbitals)~\cite{Zhang:apl,Zhang:prb22-2}, resulting in $S = 1$ or $S = 0$ states induced by Jahn-Teller distortion $Q_3$ ~\cite{Supplemental}. Hence, a $d^4$ system with a strong SOC effect, small bandwidth $W$, and cubic crystal-field splitting becomes the best candidate to obtain the $J = 0$ NM singlet insulator.

\begin{figure}
\centering
\includegraphics[width=0.48\textwidth]{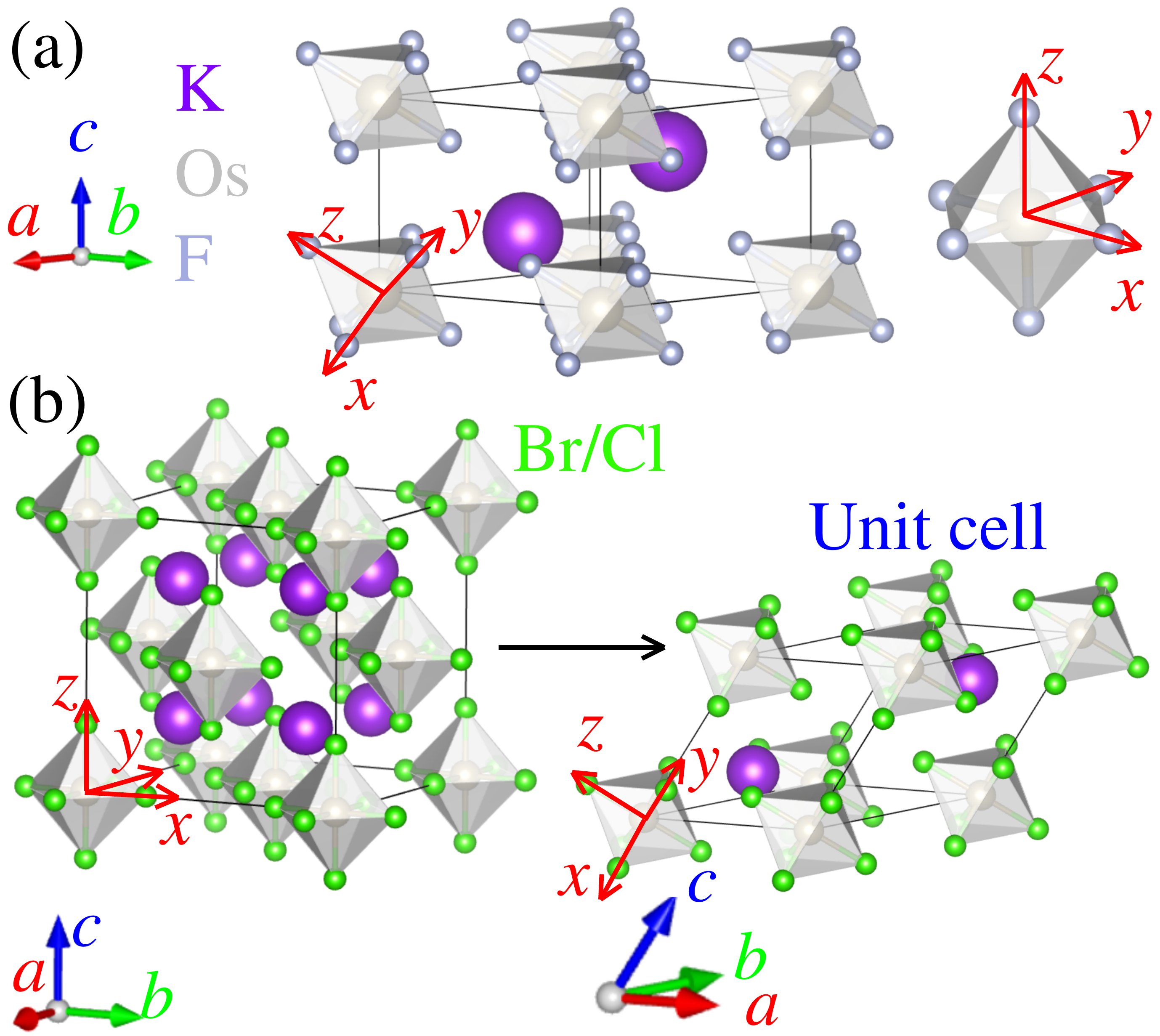}
\caption{Schematic crystal structure of K$_2$Os$X_6$ ($X$ = F, Cl and Br). (a) K$_2$OsF$_6$ with space group $P\overline{3}m1$ (No. 164). (b) K$_2$OsCl$_6$ and K$_2$OsBr$_6$ with space group $Fm\overline{3}m$ (No. 225). Crystal structures were visualized via the VESTA code~\cite{Momma:vesta}.}
\label{Fig2}
\end{figure}

K$_2$Os$X_6$ ($X$ = F, Cl, and Br) has a double-perovskite-like structure, which is known to display the required conditions. This system has a $5d^4$ electronic configuration with strong SOC Os$^{\rm 4+}$ ions, where the Os$X_6$ octahedra are at sufficiently large distance that they can be considered isolated from one another [Fig.~\ref{Fig2}]. K$_2$OsF$_6$ has the space group $P\overline{3}m1$ (No. 164) while both K$_2$OsCl$_6$ and K$_2$OsBr$_6$ have the space group $Fm\overline{3}m$ (No. 225)~\cite{crystal}. In this family, due to the well-separated Os$X_6$ octahedra, the Jahn-Teller distortion is suppressed, leading to six equal Os-$X$ bonds in this system, resulting in a nearly cubic crystal-field environment. In this case, the crystal-field splitting (between $d_{xz/yz}$ and $d_{xy}$ orbitals) is suppressed. In addition, the hopping $t$ between nearest-neighbor (NN) Os-Os sites should be small due to the Os-$X$-$X$-Os super-super exchange caused by the geometric structure of isolated Os$X_6$ octahedra. Hence, by considering the SOC in Os atoms, the $J = 0$ NM insulating state could possibly be stabilized in this family.

Based on the density functional theory (DFT) within the generalized gradient approximation (GGA) method and the Perdew-Burke-Ernzerhof revised for solids (PBEsol) exchange potential~\cite{Kresse:Prb,Kresse:Prb96,Blochl:Prb,Perdew:Prl,Perdew:Prl08}, we obtained that the relaxed crystal lattices are $a = b = 5.786$, ~and $c = 4.569$ \AA~for K$_2$OsF$_6$, close to experimental values ($a = b = 5.777$ ~and $c = 4.544$ \AA)~\cite{Ivlev:jcm}. We also found that the lattice constants of K$_2$OsCl$_6$ and K$_2$OsBr$_6$ ($a = b = c = 9.608$ \AA ~for the Cl-case, and $a = b = c = 10.184$ \AA ~for the Br-case) are in agreement with experiments ($a = b = c = 9.719$ \AA ~for the Cl-case, and $a = b = c = 10. 300$ \AA ~for the Br-case)~\cite{Takazawa:acb,McCullough:zkcm}. To save computing resources, we used the unit cell structure of K$_2$OsCl$_6$ and K$_2$OsBr$_6$ [Fig.~\ref{Fig2}(b)] in the calculations below. In addition, we also calculated the phononic dispersion, finding that these structures are dynamically stable (see Fig. S1).

\begin{figure}
\centering
\includegraphics[width=0.48\textwidth]{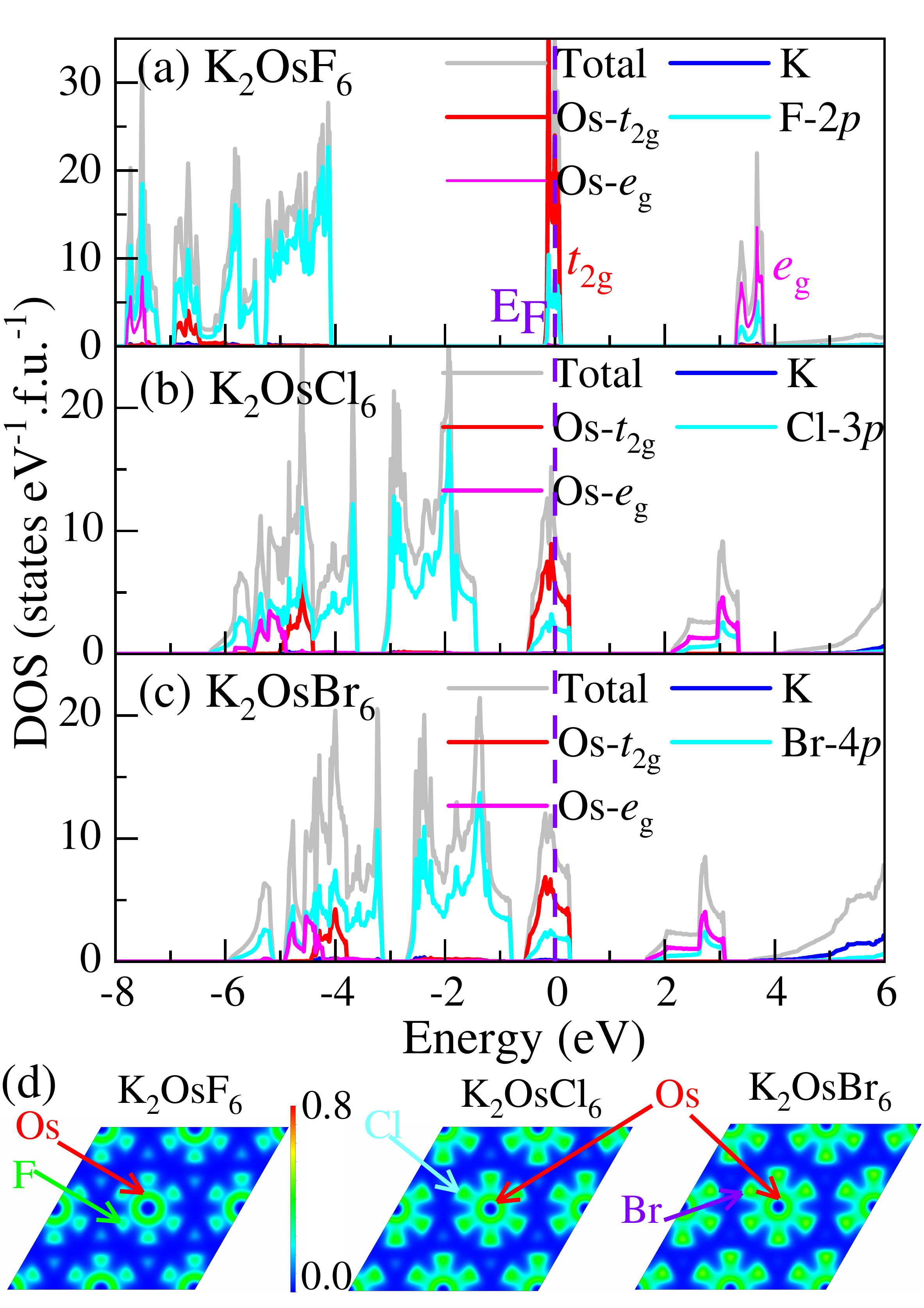}
\caption{(a)-(c) Projected density of states near the Fermi level based on the nonmagnetic states without SOC for (a) K$_2$OsF$_6$,  (b) K$_2$OsCl$_6$ and (c) K$_2$OsBr$_6$, respectively. Gray: total; red: Os; blue: K; green: F; cyan: Cl; purple: Br. The Fermi level is marked by the vertical dashed magenta line. (a) Results for the $P\overline{3}m1$ (No. 164) structure of K$_2$OsF$_6$. (b) Results for the $Fm\overline{3}m$ (No. 225) structure of K$_2$OsCl$_6$. (c) Results for the $Fm\overline{3}m$ (No. 225) structure of K$_2$OsBr$_6$. (d) Electron localization function for the $a$-$b$ plane of nonmagnetic phases of K$_2$OsF$_6$, K$_2$OsCl$_6$, and K$_2$OsBr$_6$ respectively. Generally, ELF = 0 indicates no electron localization and ELF = 1 indicates full electron localization.}
\label{Fig3}
\end{figure}

Next, we calculated the density of states (DOS) of K$_2$Os$X_6$ (X = F, Cl, and Br) in the NM state without SOC~\cite{Supplemental}. According to the DOS, the states near the Fermi level are mainly contributed by the Os-$5d$ $t_{2g}$ orbitals, partially $hybridized$ with $X$-$p$ orbitals, while most other $X's$ $p$ states are located below the Os-$5d$ energy states [Figs.~\ref{Fig3}(a-c)]. Note that the K's $4s$ states are located at high-energy bands (unoccupied states) while the K's $3p$ states occupy low-energy states below the $X$'s $3p$ states.

As shown in Figs.~\ref{Fig3}(a-c), the $X$-$p$ orbitals become closer to the Fermi level when $X$ changes from F to Br. With increasing atomic radius from F to Br, the $p$ components near the Fermi level become larger, leading to an increase in the $p-d$ hybridization tendency from F to Br. In this case, the Os's $t_{\rm 2g}$ bands are more extended in K$_2$OsCl$_6$ ($W \sim 0.7$ eV) and K$_2$OsBr$_6$ ($W \sim 0.8$ eV) than in K$_2$OsF$_6$ ($W \sim 0.3$ eV), as shown in Figs.~\ref{Fig3}(a-c), suggesting stronger electronic correlations ($U/W$) in K$_2$OsF$_6$. Furthermore, the energy splitting between the $t_{\rm 2g}$ and $e_{\rm g}$ orbitals decreases from F ($\sim 3.3$ eV) to Br ($\sim 2.6$ eV) by estimating the weight-center positions of the energy bands. In addition, we also calculated the electron localization function (ELF)~\cite{Savin:Angewandte} for K$_2$OsF$_6$, K$_2$OsCl$_6$, and K$_2$OsBr$_6$, respectively, as displayed in Fig.~\ref{Fig3}(d). The ELF picture indicates that the charges are more localized inside the Os-F bonds than the Os-Cl or Os-Cl bonds, resulting in more hybridized $p-d$ bonds in K$_2$OsF$_6$ than in K$_2$OsCl$_6$ or K$_2$OsBr$_6$. For this reason, the bandwidth of K$_2$Os$X_6$ increases from F to Br, by considering the super-super exchange coupling (Os-$X$-$X$-Os) between two NN sites of Os. Furthermore, the stronger $p-d$ hybridization tendencies would also reduce the SOC constants of Os atoms in the Cl- or Br-cases. This reduction was also experimentally observed in K$_2$Ir$X_6$ ($X$ = F, Cl and Br) with the $5d^5$ configuration ($572/444/420$ meV for F/Cl/Br, respectively)~\cite{Reig-i-Plessis:Prm}.

To understand qualitatively the possible $J = 0$ state based on the maximally localized Wannier functions (MLWFs) method~\cite{Mostofi:cpc}, we obtained the on-site energies and hoppings for different Os's $t_{\rm 2g}$ orbitals~\cite{wannier}. The spreads of $t_{\rm 2g}$ orbitals of K$_2$OsF$_6$ are much smaller than that of K$_2$OsCl$_6$ and K$_2$OsBr$_6$, indicating a more localized behavior in K$_2$OsF$_6$ than in the Cl- or Br- cases. Furthermore, the $d_{xy}$, $d_{yz}$, and $d_{xz}$ orbitals have almost the same on-site energies, indicating that the crystal-splitting energy is nearly zero,  thus achieving the condition needed for a stable $J = 0$ state. Moreover, the largest elements of the hopping matrix of NN Os-Os sites are 40, 88, and 95 meV for K$_2$OsF$_6$, K$_2$OsCl$_6$, and K$_2$OsBr$_6$, respectively. Based on the values of hoppings,  the on-site energies of the $t_{\rm 2g}$ orbitals ($\Delta$ $\sim 0$), using typical electronic correlations of Os atoms ($U$ $\sim 1-2$ eV, $J_H$ $\sim 0.3-0.4$ eV), and the strong SOC of the Os atom ($\sim 0.4$ eV)~\cite{Rau:arcm,Takayama:jpsj,Zhang:apl,Zhang:prb22-2}, the $J = 0$ NM state should be obtained, as discussed in Hubbard model studies~\cite{Khaliullin:prl13,Meetei:prb15,Kaushal:prb17}.  In the limit of large on-site Hubbard coupling $U$
$\gg$ $\lambda$, naively the system would be an $S = 1$ Mott state where the electronic correlations play the dominant role. However, this is not the case we studied here. In addition, our GGA+$U$+SOC calculations also provide a NM ground state for all F-, Cl- and Br-cases~\cite{Supplemental}.

\begin{figure*}
\centering
\includegraphics[width=0.96\textwidth]{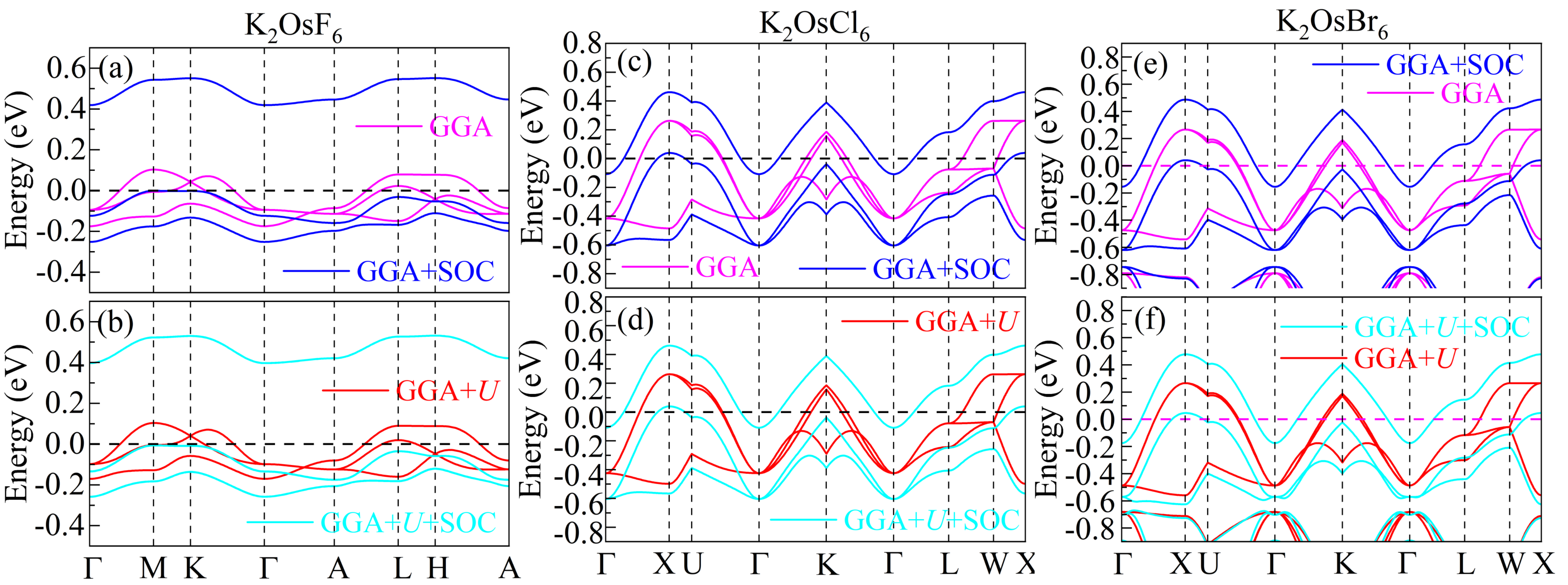}
\caption{Band structures of the NM state based on GGA, GGA+SOC, and GGA+$U$+SOC for (a-b) K$_2$OsF$_6$, (c-d) K$_2$OsCl$_6$, and (e-f) K$_2$OsBr$_6$, respectively. Here, the correlation effects were considered by the Liechtenstein formulation within the double-counting term~\cite{Liechtenstein:prb}, where the on-site Coulomb interaction used was $U = 2$ eV, and the Hund coupling was $J_H = 0.4$~eV for the Os atoms~\cite{J/U}. The Fermi level is the horizontal dashed line. (a-b) For K$_2$OsF$_6$, the coordinates of the high-symmetry points in the Brillouin zone (BZ) are given by $\Gamma$ = (0, 0, 0), M = (0.5, 0, 0), K = (1/3, 1/3, 0), A = (0, 0, 0.5), L = (0.5, 0, 0.5) and H = (1/3, 1/3, 0.5). (c-f) For K$_2$OsCl$_6$ or K$_2$OsBr$_6$, the coordinates of the high-symmetry points in the BZ are given by $\Gamma$ = (0, 0, 0), X = (0.5, 0, 0.5), U = (0.625, 0.25, 0.625), K = (0.375, 0.375, 0.75), L = (0.5, 0.5, 0.5) and W = (0.5, 0.25, 0.75).}
\label{Fig4}
\end{figure*}

To better understand the possible $J = 0$ NM insulating state, we calculated the band structures with/without SOC or $U$ for K$_2$OsF$_6$, K$_2$OsCl$_6$, and K$_2$OsBr$_6$, respectively, as displayed in Fig.~\ref{Fig4}. First, let us focus on discussing the electronic structures for K$_2$OsF$_6$. Without SOC and $U$ effects, Fig.~\ref{Fig4}(a) displays a strong metallic behavior since the $t_{\rm 2g}$ of Os states are not completely occupied (four electrons occupy three $t_{\rm 2g}$ orbitals). By introducing the SOC effect, the $t_{\rm 2g}$ orbitals of K$_2$OsF$_6$ are divided into $j_{\rm eff} = 1/2$ and $j_{\rm eff} = 3/2$ states, separated by an energy gap between the two $j_{\rm eff}$ states, as shown in Fig.~\ref{Fig4}(a). Then, four electrons of the Os$^{\rm 4+}$ ($5d^4$ configuration) fully occupy the lower $j_{\rm eff} = 3/2$ quadruplet, leading to an unoccupied $j_{\rm eff} = 1/2$ doublet, resulting in a $J = 0$ NM insulator.

As shown in Fig.~\ref{Fig4}(b), we also studied the band structures with the electronic correlations included~\cite{U}. By only introducing the electronic correlation $U$, the band structure of the $t_{\rm 2g}$ states of K$_2$OsF$_6$ is similar to the GGA case, where the $t_{\rm 2g}$ bands of Os are not separated and keep metallic behavior. By considering both the SOC and $U$ effects, the $t_{\rm 2g}$ of K$_2$OsF$_6$ orbitals are reconstructed to the $j_{\rm eff} = 1/2$ and $j_{\rm eff} = 3/2$ states, opening a gap. In this case, the SOC plays the key role in deciding the nature of the insulating state, by separating the empty $j_{\rm eff} = 1/2$ and fully-occupied $j_{\rm eff} = 3/2$ states. The almost undistorted Os$X_6$ ($X$ = F, Cl, and Br) octahedra are ideally separated, leading to a dramatic decrease in the hopping between Os-Os sites. In this quasi-disconnected geometry, the weak Os-$X$-$X$-Os superexchange interaction leads to the decreasing connectivity of the Os$X_6$ octahedra, resulting in a case close to the atomic limit. In the F-case, the large SOC effect achieves a $J = 0$ NM state by comparison with the small hopping ($\sim 40$ meV) and quenched crystal-field splitting.

Without SOC and $U$ effects, the band structures of K$_2$OsCl$_6$ and K$_2$OsBr$_6$ show metallic behavior due to the  partially occupied $t_{\rm 2g}$ orbitals, as shown in Figs.~\ref{Fig4}(c) and (e). By introducing the SOC in K$_2$OsCl$_6$ and K$_2$OsBr$_6$, the $t_{\rm 2g}$ bands begin to separate and reconstruct into the $j_{\rm eff} = 1/2$ and $j_{\rm eff} = 3/2$ states [see Figs.~\ref{Fig4}(c) and (e)]. Different to the results for K$_2$OsF$_6$, a band gap is not obtained with the SOC effect in both K$_2$OsCl$_6$ and K$_2$OsBr$_6$, keeping a metallic state. Some Os $5d$ electrons partially occupy the $j_{\rm eff} = 1/2$ states, contributing to the conductivity in both the Cl- and Br- cases. This can be understood intuitively. Due to the stronger $p-d$ hybridization tendencies in the Cl- or Br-cases than in the F-case, the bandwidth increases, and the SOC constants of Os atoms are reduced. Then, the hopping $t$ could compete with the strong SOC $\lambda$, leading to a metallic phase with partially occupied $j_{\rm eff} = 1/2$ states crossing the Fermi level. In this region, the $j_{\rm eff} = 1/2$ and $j_{\rm eff} = 3/2$ states are not totally separated. Then, if the strength of the SOC could continues to increase, the gap should finally open. Furthermore, similar to the results for K$_2$OsF$_6$, the electronic correlation $U$ would not separate the $t_{\rm 2g}$ states and open a gap [Figs.~\ref{Fig4}(d) and (f)].

\begin{figure}
\centering
\includegraphics[width=0.48\textwidth]{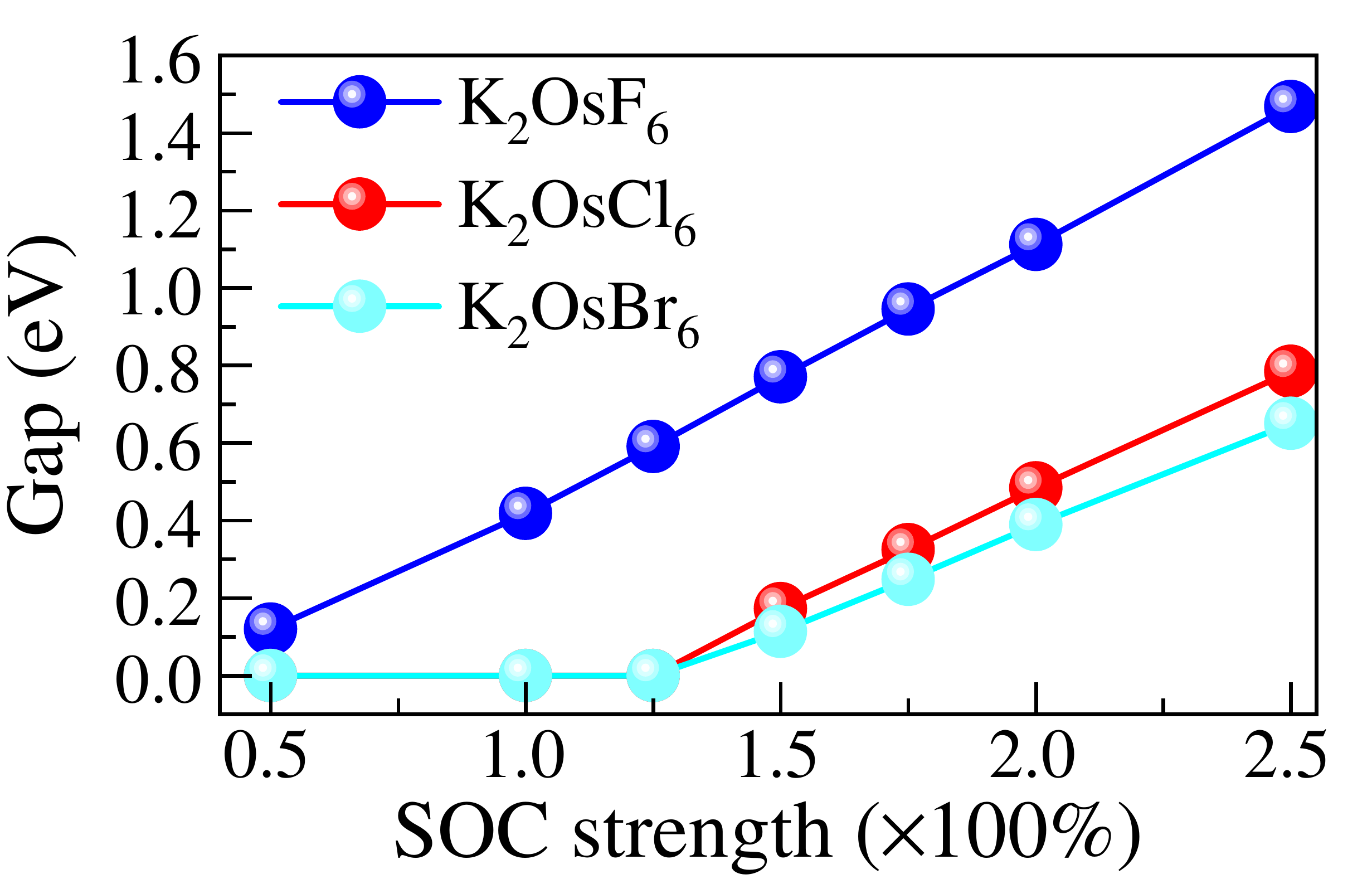}
\caption{ Band gaps as a function of SOC strength for K$_2$OsF$_6$, K$_2$OsCl$_6$, and K$_2$OsBr$_6$, respectively, within GGA+SOC calculations. Here, we do not introduce the Hubbard interaction $U$ on Os sites because the SOC plays the key role to open the gap.}
\label{Fig5}
\end{figure}

To better understand the SOC effect, we also calculated the band gaps within the GGA+SOC approximation by selecting various SOC strengths. As displayed in Fig.~\ref{Fig5}, the gap increases by enhancing the SOC strength~\cite{SOC-change}. Due to the small bandwidth of K$_2$OsF$_6$, the gap can be opened by a small SOC strength ($\sim 50\%$). On the other hand, the insulating gap can be obtained for a higher SOC strength ($\sim 150\%$) for K$_2$OsCl$_6$ and K$_2$OsBr$_6$ because the larger bandwidths of the Cl- or Br- cases leads to a competition between hopping $t$ and SOC $\lambda$. Because the energy gap between the Os's $j_{\rm eff}$ = 1/2 and $j_{\rm eff}$ = 3/2 states is about $3\lambda /2$, we can estimate the SOC values by calculating the changes in the value of this energy gap under very small modifications in the SOC strength. In addition, we also estimated that the SOC values are about $0.467$, $0.405$, and $0.367$ eV for K$_2$OsF$_6$, K$_2$OsCl$_6$, and K$_2$OsBr$_6$, respectively. This reduced tendency of SOC values in K$_2$Os$X_6$ is quite similar to the case of the K$_2$Ir$X_6$ system with $d^5$ configuration~\cite{Reig-i-Plessis:Prm}. By comparing the induced hoppings from F to Cl, this could explain the metallic behavior in Cl- and Br- cases. By continuing increasing the strength of the SOC, the gaps increase in K$_2$OsCl$_6$ and K$_2$OsBr$_6$, as expected. In the strong SOC condition, the entire K$_2$Os$X_6$ (X = F, Cl, and Br) family could be considered to be a potential $J = 0$ NM insulator. The insulating vs. metallic character of the correlated system under study is decided by several factors related to different parameters (primarily $t$, $\lambda$, $J_H$, and $U$). As described in the text, the $J = 0$ state can be stabilized in a $d^4$ system only in a special range of parameters, namely under severe limitations, at least according to our calculations. For a complete physical picture, it would be important to study how the $J = 0$ state evolves by varying those many different parameters. Such effort will demand considerable computational resources and discussion, and they should be based on model calculations, beyond the scope of our present manuscript.

For the spin-orbital-entangled $J = 0$ NM insulating compounds, the excitonic $J = 1$ triplet state displays interesting magnetic order caused by the condensation of mobile spin-orbit excitons~\cite{Khaliullin:prl13,Meetei:prb15}. The ordered moment is dependent on the competition between exchange interactions and the energy gap caused by the SOC. Furthermore, some possible interesting features can be obtained near the quantum critical point (QCP) in the $J = 1$ excitonic state, such as a Higgs mode~\cite{Jain:np} and magnons~\cite{Takayama:jpsj}. Our results for the K$_2$Os$X_6$ (X = F, Cl, and Br) family provide a starting point for  experimentalists and theorists to work on the $J_{\rm eff} = 0$ state or the $J = 1$ triplet excitations on this $5d^4$ system, such as in inelastic neutron scattering(INS) or resonant inelastic x-ray scattering (RIXS) experiments. In fact, the K site could be replaced by other $1+$ ions~\cite{Armstrong:pr}, such as Rb$^+$, Cs$^+$ and (NH$_4$)$^+$, where the same $J = 0$ physics should be obtained. Due to the ``zero-dimensional" geometry structure, the K site may be replaced by $2+$ ions, leading to a $d^5$ configuration, where the spin-orbital Mott state could be obtained. Hence, our results clearly provide a potential candidate system for experimentalists and theorists to work on this K-system and related materials.

In summary, we presented a systematic study of the K$_2$Os$X_6$ ($X$ = F, Cl, and Br) family with a $5d^4$ electronic configuration by using DFT first-principles calculations. Due to the isolated geometry of the well-separated Os$X_6$ octahedra, this system is close to the cubic crystal-field limit and results in dramatically decreasing hoppings for nearest-neighbor Os-Os sites, providing a fertile condition for obtaining the $J = 0$ NM insulator. By introducing the SOC, the three degenerate $t_{2g}$ orbitals are split into two separated ``effective'' $j_{\rm eff}$ ($j_{\rm eff} = 1/2$ and $j_{\rm eff} = 3/2$ states) states. In K$_2$OsF$_6$, due to the small bandwidth of the Os $5d$ orbitals ($\sim 0.3$ eV), the SOC effect is sufficiently strong to open a gap. Hence, four electrons of the Os$^{\rm 4+}$ ($5d^4$ configuration) fully occupy the lower $j_{\rm eff} = 3/2$ quadruplet, leading to an unoccupied $j_{\rm eff} = 1/2$ doublet, resulting in a $J = 0$ NM insulator.

Furthermore, the hybridization between the Os $5d$ orbitals and $X$ ($X$ = F, Cl, and Br) $p$ orbitals increases from F to Br, leading the electrons in K$_2$OsCl$_6$ and K$_2$OsBr$_6$ to be less localized than in K$_2$OsF$_6$, resulting in a larger bandwidth for the Cl- or Br- cases ($\sim 0.7/0.8$ eV for K$_2$OsCl$_6$ and K$_2$OsBr$_6$, respectively) than in the F case. In these compounds, the SOC $\lambda$ competes with the hopping $t$ in K$_2$OsCl$_6$ and K$_2$OsBr$_6$, and the combination is not enough to open a gap because some electrons would occupy the $j_{\rm eff} = 1/2$ states. By increasing the SOC strength to $\sim 150 \%$, the $j_{\rm eff} = 1/2$ and $j_{\rm eff} = 3/2$ states become totally separated, obtaining the $J = 0$ NM insulator in K$_2$OsCl$_6$ and K$_2$OsBr$_6$. Hence, our results should encourage experimentalists and theorists to continue working on this interesting family of osmium halides to achieve the $J_{\rm eff} = 0$ state, and  also $J = 1$ triplet excitations and excitonic magnetism.

The work of Y.Z., L.-F.L., A.M. and E.D. is supported by the U.S. Department of Energy (DOE), Office of Science, Basic Energy Sciences (BES), Materials Sciences and Engineering Division.


\begin{references}
\bibitem{crystal} K$_2$OsF$_6$ has a trigonal space group (No. 164), where the OsF$_6$ octahedra have a slightly distorted octahedron-like behavior, resulting in the angle of the F-Os-F bond not being strictly $90^\circ$. In K$_2$OsCl$_6$ and K$_2$OBr$_6$, the OsF$_6$ or  OsBr$_6$ octahedra are ideal without any distortion, leading to the ideal $90^\circ$ bonds in the cubic structure (No. 225).
\bibitem{wannier} Here, we constructed three disentangled Wannier functions based on the MLWFs method~\cite{Mostofi:cpc} to fit the low-energy DFT bands near the Fermi level, involving the $t_{2g}$ orbital basis ($d_{xy}$, $d_{yz}$, and $d_{xz}$) for each Os atom in the nonmagnetic phase without SOC.  The local basis is marked in the inset of Fig.~\ref{Fig1}, with the $x$-, $y$-, or $z$-axis along with the Os-$X$ directions. As shown in Fig. S2, the DFT bands near the Fermi level are fitted very well with the Wannier bands obtained from MLWFs. Due to a slightly distorted octahedron-like behavior in K$_2$OF$_6$, the calculated on-site energies of $t_{\rm 2g}$ orbitals are slightly different ($\sim 0.6$) meV, approaching the cubic filled condition. Due to the undistorted octahedra in K$_2$OCl$_6$ and K$_2$OBr$_6$, we obtained the same on-site energy for the three degenerate $t_{\rm 2g}$ orbitals.
\bibitem{SOC-change} Here, $100\%$ SOC means the standard SOC calculations in the original DFT code. Then, we change different strength ratios for the SOC to study the electronic structures based on GGA+SOC calculations, i.e. $200\%$ indicates two times of original SOC strength.\
\end{references}
\end{document}